# Submesoscale circulation in the Northern Gulf of Mexico:

# Deep phenomena and dispersion over the continental slope

Annalisa Bracco[1*], Keshav Joshi[1,2], Hao Luo[1,3], James C. McWilliams[4]

[1] School of Earth and Atmospherics Sciences, Georgia Institute of Technology, Atlanta, GA 30332, USA

[2] School of Physics, Georgia Institute of Technology, Atlanta, GA 30332, USA

[3] Department of Marine Sciences, University of Georgia, Athens, GA 30602, USA

[4] Department of Atmospheric and Oceanic Sciences, University of California, Los Angeles, CA 90095, USA

Corresponding Author:

Annalisa Bracco, School of Earth and Atmospheric Sciences,

311 Ferst Dr, Atlanta, GA, 30332, USA

email: abracco@gatech.edu

Ph: 404-894-1749

**Abstract**

This study examines the mesoscale and submesoscale circulations along the continental slope in the northern Gulf of Mexico at depths greater than 1000 m. The investigation is performed using a regional model run at two horizontal grid resolutions, 5 km and 1.6 km, over a three year period from January 2010 to December 2012.

Ageostrophic submesoscale eddies, and vorticity filaments populate the continental slope, and they are stronger and more abundant in the simulation at higher resolution, as to be expected. They are formed from horizontal shear layers at the edges of highly intermittent, bottom-intensified, along-slope boundary currents and in the cores of those currents when they are confined on steep slopes. The horizontal resolution influences the representation of the lateral and vertical transport of neutrally buoyant tracers. This is exemplified by a passive dye deployed near the Macondo Prospect, where the *Deepwater Horizon* rig exploded in 2010, and at the largest known natural hydrocarbon seep in the northern Gulf, known as GC600, located few hundred kilometers to the west of the rig wellhead. The first site is characterized by weak mean currents and, in the high resolution run, by intense submesoscale eddies that form to the west of the Macondo Prospect near the ocean bottom. GC600, on the other hand, is surrounded by a complicated bathymetry, and its flow is strong and highly variable and, in the higher resolution case, characterized by short lived submesoscale eddies. The trapping exerted by the submesoscale eddies and vorticity filaments near the rig constrain the spreading in the vicinity of the Macondo Prospect, and as a result, patchiness in the passive tracer distributions differs in the two integrations, despite similar diffusivities expressing the average spreading rate. The tracer released at GC600, on the other hand, is dispersed more effectively in the first six months in the higher resolution case due to enhanced near-bottom turbulent mixing.

## 1. Introduction

Processes and dynamics occurring at the oceanic submesoscales (horizontal length scales between ~ 100 m and ~ 10 km, and time scales between a few hours and a few days) have been shown to impact the stirring and dispersion of natural and anthropogenic tracers in the upper ocean. Submesoscale circulation patterns can modify the mixed-layer stratification, and manifest as ageostrophic fronts and, whenever these fronts are unstable, meanders and eddies (Capet et al., 2008b; Thomas et al., 2008; McWilliams et al., 2009a; Taylor and Ferrari, 2011). The generation of submesoscale fronts and consequent instabilities has been investigated in a number of papers (Capet et al., 2008a; McWilliams et al., 2009b; Thomas and Ferrari, 2008).

A heuristic description of front generation that is relevant to this work is presented in Lévy et al. (2012): submesoscale ageostrophic frontogenesis can take place whenever density gradients steep enough that their Rossby number, $R_o$, defined as $U/fL$ (where $U$ is the characteristic flow velocity, $f$ is the Coriolis frequency, and $L$ is the width of the flow pattern) is O(1). In the oceanic interior any density anomaly generated by the stirring and straining of the mesoscale field is quickly damped by overturning circulations that reduce the local steepness in the density field to restore geostrophic balance, with the end result that almost flat density surfaces are maintained. At the top and bottom of the ocean, however, those overturning circulations are limited by the presence of the air-sea and bottom boundaries, and must develop along horizontal surfaces, with the end result of contributing to further sharpening the density fronts, and therefore to frontogenesis.

Several recent studies attempted to characterize submesoscale dynamics near the ocean surface (Boccaletti et al., 2007; Capet et al., 2008a,b; Thomas et al., 2008; Thomas and Ferrari, 2008) and to quantify their role on the transport and mixing of tracers with numerical tools in both idealized and realistic domains (Klein and Lapeyre, 2009; Koszalka et al., 2009; Lévy et al., 2010; Lévy et al., 2011; Zhong et al., 2012; Zhong and Bracco, 2013; Gula et al., 2014) and with targeted field campaigns (D'Asaro et al., 2011; Poje et al., 2014; Shcherbina et al., 2013).

Little is known, however, about those dynamics and their effects at depth, near the sea floor, with the exception of the California Current System, where ageostrophic submesoscale motions have been shown to contribute to the generation of long-lived subsurface anticyclones through centrifugal instability (Molemaker et al., 2015; Dewar et al., 2015). In this work we investigate the impact of submesoscale dynamics along the continental slope in the northern Gulf of Mexico (hereafter GoM).

The GoM basin is characterized by a broad, highly variable slope that hosts a large number of natural hydrocabon seeps (Peccini and MacDonald, 2008; Garcia-Pineda et al., 2009) found predominantly at depths greater than 1000 m. It was impacted by the largest deep-water oil spill ever recorded following the explosion of the *Deepwater Horizon* rig on April 20, 2010. The spill discharged approximately $3 \times 10^8$ kg of gas and between 6 and $8 \times 10^8$ kg of oil, about a third of which was confined in underwater plumes found in several layers between 700 m and the ocean seafloor (Joye et al., 2011).

In this paper we address two main questions with the goal of improving our understanding of transport and mixing along continental slopes:

- Are submesoscale dynamics relevant to the circulation along the continental slope in the northern Gulf of Mexico, and if so what are the mechanisms responsible for their generation?
- Which role, if any, does model resolution play in the representation of transport and mixing of passive, neutrally buoyant tracers released along the slope and in particular in the vicinity of the *Deepwater Horizon* site?

We consider two simulations of the northern Gulf of Mexico, performed with 5 km and 1.6 km horizontal grid resolutions and spanning three years, from January 2010 to December 2012; the first resolves fully the mesoscale dynamics, while the second one is submesoscale permitting (i.e., has partial resolution). We concentrate on dynamics relevant for the deep layer in the GoM, considering only depths approximately equal or greater than 1000 m. After analyzing the vorticity structures that characterize the flow in the two runs, we present how passive tracers deployed at two locations evolve over one year, and quantify the tracer diffusivities. While the simulations we consider are probably not directly relevant to the 2010 oil spill --- particular realizations of oceanic eddies are inherently unpredictable without observational constraints --- the answers to the questions above may guide better hindcasting and forecasting capabilities for deep pollutant releases.

## 2. Model setup, domain and forcing fields

The Gulf of Mexico circulation is simulated with the Regional Oceanic Modeling System (ROMS) in the version developed by the Institut de Recherche pour le Développement (IRD), ROMS-Agrif 2.2 (Debreu et al., 2012). The model domain covers the whole Gulf of Mexico (GoM), but we concentrate on its northern portion and specifically on the area comprised between 96.31°W - 86.93°W and 25.40°N - 30.66°N, indicated as northern Gulf in the following (Fig. 1). This region contains a large number of natural hydrocarbon seeps (MacDonald et al., 2002) and coincides with the area that was the most affected by the *Deepwater Horizon* spill in 2010 (Joye et al., 2011). In this work we consider two integrations differing only in their horizontal grid resolutions of 5 km (LR, for Low Resolution) and 1.6 km (HR, for High Resolution). HR is obtained exploiting the two-way nesting capabilities of ROMS-Agrif 2.2 in the focus area. The vertical resolution is 70 terrain-following layers enhanced near the surface and bottom. The model bathymetry is derived from the 2 min Gridded Global Relief Data Collection topography ETOPO2 (Sandwell and Smith, 1997), interpolated at 5 km and smoothed with a maximum slope parameter of 0.35 (Penven et al., 2008) to avoid spurious currents generated by pressure gradient errors. In the nested solution the bathymetry is interpolated from its 5 km version, so that no new, small-scale topographic features are added. Both runs span three years, from January 2010 to December 2012, and they are identically forced by 6-hourly momentum and 12-hourly heat fluxes from the ERA-Interim reanalysis (Dee et al., 2011). The open ocean boundaries are nudged to the monthly fields from the HYCOM – NCODA (Hybrid Coordinate Ocean Model - Navy Coupled Ocean Data Assimilation) ocean prediction system (GOMI0.04 expt. 30.1, http://www7320.nrlssc.navy.mil/hycomGOM). Fresh water fluxes are prescribed by nudging the surface salinity field to the World Ocean Atlas 2009 (WOA09) monthly varying climatology (Antonov et al., 2010) without any synoptic or interannual variability. More information about the model configuration, spin-up, and validation can be found in Luo et al. (2015) (LBCM in the following). Configuration and forcing files are also available through the Gulf of Mexico Research Initiative Information and Data Cooperative (GRIIDC) under number: R1.x132.141:0005.

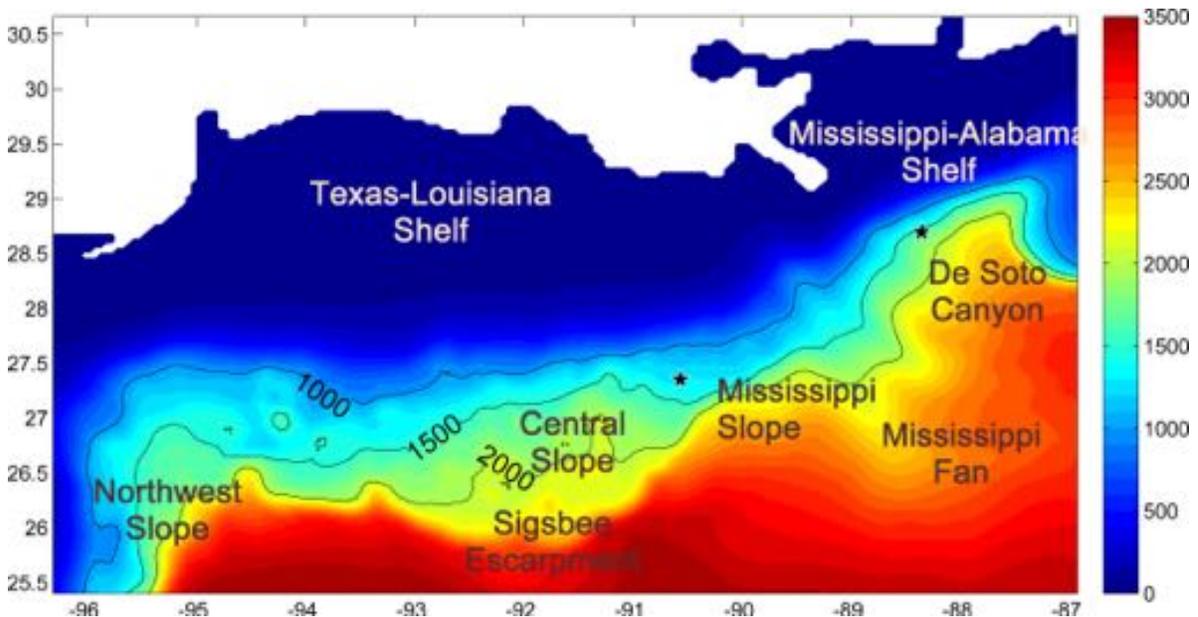

**Figure 1.** Bathymetry over the northern Gulf (HR) domain. The 1000, 1500, and 2000 m bathymetric contours are indicated by the black lines and major topographic features are named. The sites where the passive tracer is released are indicated by black stars.

In both integrations passive tracers are deployed at a site 6 km to the south-southeast of Mississippi Canyon Block 252 (abbreviated MC252), where the *Deepwater Horizon* oil drilling rig was located, MC297 (-88.36°W, 28.70°N, depth 1562 m). A second release is carried out at Green Canyon Block 600 (GC600, -90.57°W, 27.36°N, depth 1249.3 m), the largest natural oil and gas seep in the northern Gulf located at roughly the same water depth of MC252 but to the west of the Mississippi Fan, along the eastern edge of the Central Slope. Both deployment sites have been visited regularly by cruises of the ECOGIG consortium (https://ecogig.org) between 2010 and 2013. The tracers are released on April 20, 2010 and are advected for 12 months, using a split tracer advection and diffusion scheme and the rotated biharmonic diffusion with flow-dependent hyperdiffusivity that satisfies the Peclet constraint (Marchesiello et al., 2009; Lemarié et al., 2012).

## 3. Mean Circulation at depth

The circulation in the Gulf can be approximated as a two-layer system. The top layer extends approximately over the upper 1000 m and its circulation is dominated by the anticyclonic Loop Current and the Rings or Loop Eddies that detach from it on average every ten-eleven months (Vukovich, 2007), together with associated pycnocline mesoscale and surface-layer submesoscale currents. Underneath, the mean flow is cyclonic around the basin away from the continental slope, as indicated by current-meter observations, Lagrangian circulation (PALACE) floats, and the temperature structure in the basin (DeHaan and Sturges, 2005; Hamilton, 1990; Weatherly, 2004). In-situ measurements reveal that vortex stretching and topographic Rossby waves (TRWs) with periodicities ranging from 10 to 50 days, contribute to the instantaneous deep flow measured in the modeled portion of the northern Gulf (DeHaan and Sturges, 2005; Hamilton, 2009; Kolodziejczyk et al., 2012; Pérez-Brunius et al., 2013). TRWs in particular are responsible for the bottom intensification of currents aligned and trapped along the continental slope where the bathymetry is steep (Dukhovskoy et al., 2009; Hamilton, 2007; Hamilton and Lugo-Fernandez, 2001).

Model simulations confirm the mean flow direction (Lee and Mellor, 2003; Cardona and Bracco, 2014) and the approximate periodicity of the intra-seasonal variability

(see Appendix). Fig. 2 shows the modeled mean velocities at 1500 m depth superimposed on the mean speed, $\sqrt{\bar{u}^2 + \bar{v}^2}$, where $\bar{u}(x,y)$ and $\bar{v}(x,y)$ are the horizontal velocity components time-averaged over the integration period. A cyclonic current dominates the circulation south of 26°N. Over the slope the mean flow consists of numerous closed recirculations generated by the interaction with the complex bathymetry, with similar patterns in the two integrations. The mean speed at 1500 m in Fig. 2, while representing only a small portion of the total kinetic energy of the flow, is about 35% higher in HR than in LR. The difference in mean speeds should not be attributed to resolution, but is more an expression of the intrinsic variability of the system, likely linked to the different realizations of the upper layer dynamics in relation to the modeled LC variability[1]. The cyclonic current pattern south of 26.3°N over the deeper portion of the basin is the most prominent feature in the mean flow independent of resolution, and it is a major contributor to the HR versus LR mean speed amplitude difference. This current is coherent throughout the deep layer in agreement with the analysis of trajectories from 36 RAFOS floats deployed for a year or less between April 2003 and April 2004 at depths between 1000 and 3000 m (Hamilton, 2009). Additionally, its amplitude increases with depth, both in the model and in the observations.

Despite the identical atmospheric forcing and boundary conditions, the mesoscale dynamics and in particular the Loop Current evolution are not deterministic (Cardona and Bracco, 2014), and the outcomes differ in the two runs (LBCM); note that no observational data are assimilated. The domain averaged eddy kinetic energy (EKE) differs between HR and LR, and it is larger in HR. The slightly greater amplitude of the instantaneous velocity field in HR usually compares better with the (limited) available observations (DeHaan and Sturges, 2005, Hamilton and Lugo-Fernández, 2001, Oey and Hamilton, 2012; Weatherly, 2004). Modeled near-bottom velocities are further validated in the Appendix.

---

[1] We have verified that the mean speed of a different 8-year long run at 5 km resolution presented in Cardona and Bracco (2014) is indeed comparable to HR in both patterns and intensity.

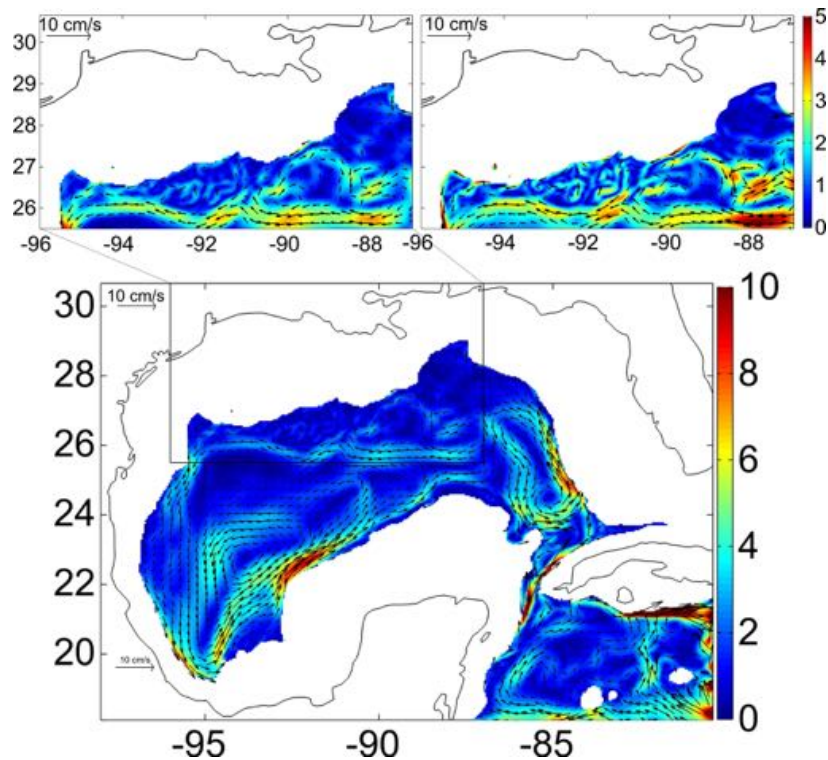

**Figure 2.** Mean velocity vector field in the whole Gulf of Mexico with zooms in the northern portion in both HR (top left) and LR (top right). The color bar indicates speed in cm s$^{-1}$.

EKE time series at various depths, calculated as $\frac{1}{2}(\langle u'^2 + v'^2 \rangle)$, where <> designates spatial averaging over the northern Gulf (96.31°W - 86.93°W and 25.40°N - 30.66°N) and the mean velocity components at each grid point have been subtracted, are shown in Figure 3. The time series highlight the increase of EKE with depth, its coherency through the deep layer, and its intensification at irregular intervals between 2-7 months in both runs. The EKE peaks have comparable amplitudes in HR and LR, and they are associated with the intermittent strengthening of the along-slope current system contributing to the recirculation region centered at -89.5°W, and with the formation of a mesoscale cyclone as discussed in the next section. Large velocity fluctuations with a time-scale comparable to the modeled ones have been recorded in the same area (Hamilton and Lugo-Fernández, 2001), but their generation mechanism remains unexplained. As noted in Cardona and Bracco (2014), the deep Gulf circulation does not exhibit any seasonality, and the EKE variability below 1000 m is uncorrelated to the surface EKE (see Figure A.1 in LBCM for the surface EKE time series in both HR and LR runs).

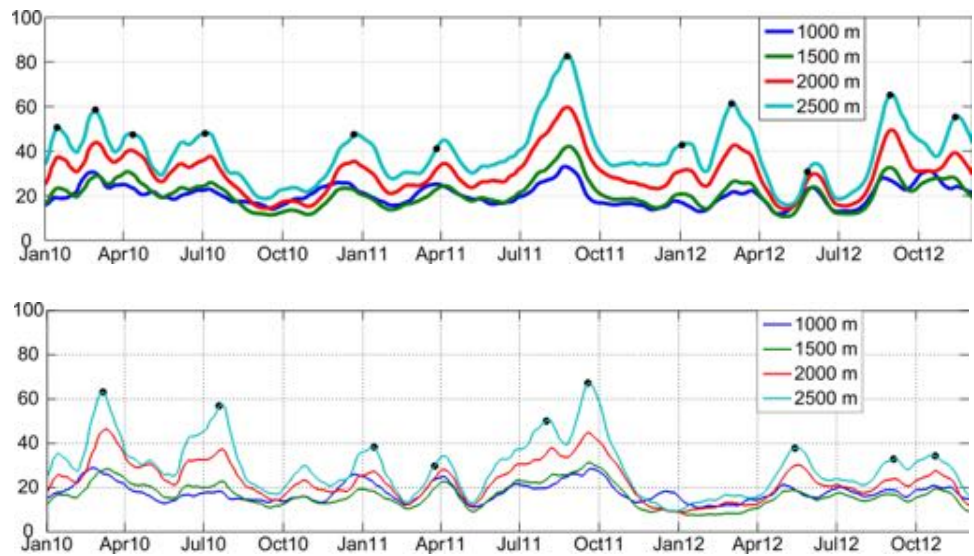

**Figure 3.** Eddy Kinetic Energy time series averaged over the northern Gulf at different depths across the deep layer. HR run on top (thick lines) and LR integration below

(thin lines). The black dots indicate the generation of mesoscale cyclones at the Mississippi Fan. Unit: cm²s$^{-2}$

## 4. Mesoscale and submesoscale dynamics at depth

Two typical snapshots of relative vorticity normalized by the Coriolis parameter, $\zeta/f$, depth-averaged between 1200 and the ocean bottom, are shown in Fig. 4 (September 2011 in HR and August 2010 in LR). Both snapshots include an occurrence of the largest, strongest ($\zeta \geq 0.8\,f$) feature found in the deep layer in the northern Gulf: a bottom intensified mesoscale cyclone that forms at the Mississippi Fan. The episodic generation of this cyclonic eddy is indicated by the black dots in the EKE time series.

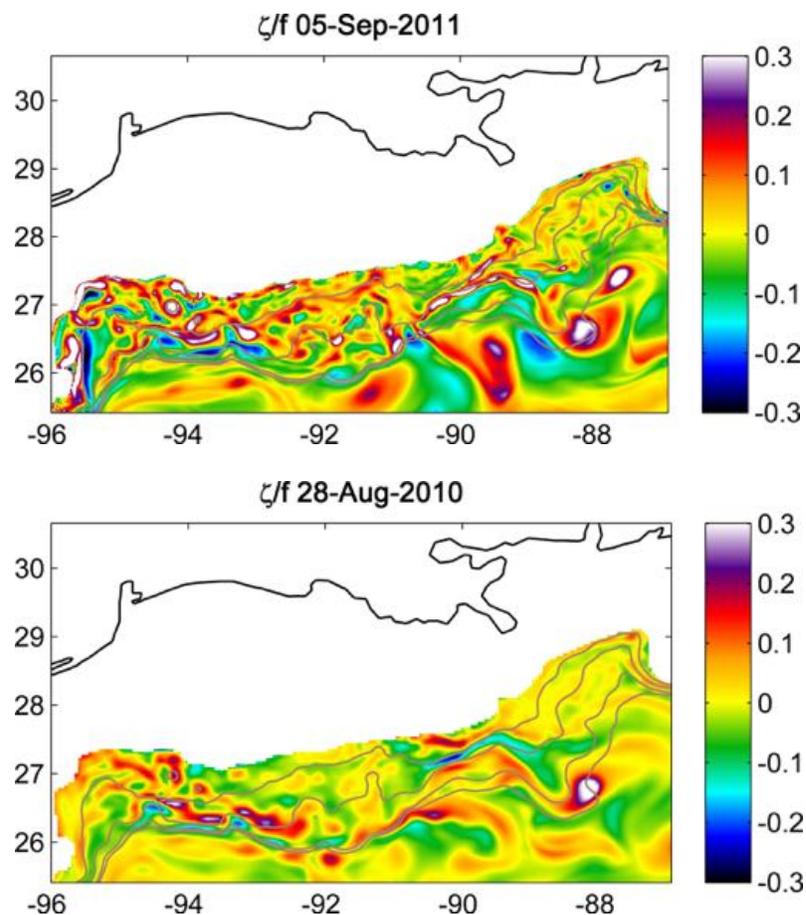

**Figure 4.** Snapshots of relative vorticity normalized by the Coriolis frequency, $\zeta/f$, depth-averaged from 1000 m to the bottom. Top panel: HR; Bottom panel: LR.

The cyclone formed at the Fan is coherent throughout the lower layer and bottom-intensified, as indicated by the synchronous peaks in the EKE time-series at different depths in Fig. 3. The vorticity generation at the Fan results from the interaction of the temporally varying current and the bathymetry. Analogous cyclogenesis has been detected using bottom current meters in the Kuroshio Extension (Greene et al., 2009). The along slope lower-layer flow to the west of the Fan is generally advected eastward contributing to the recirculation zone centered around -89.5ºW in Fig. 2. If the zonal velocity is weak, the flow is inhibited from moving over the seamount, but if the along-slope current is strong enough, the flow is advected over and around the topographic feature, where vortex lines are compressed and anticyclonic vorticity is generated as a consequence of potential vorticity conservation, as in the case of a uniform flow impinging on a seamount modeled in Huppert and Bryan (1976). In our runs, a portion of the flow is advected around the Fan, decreasing somewhat the anticyclonic vorticity induced atop the Fan. At the same time, for sufficiently strong currents, the fluid atop the Fan is advected eastward and cyclonic vorticity results from vortex stretching (Fig. 5, top panel). The anticyclonic vorticity is trapped on the Fan, while the cyclone can drift downstream (Fig. 5, middle and bottom panels) and away from the topographic feature until it is entrained in the mean cyclonic current south of 26.3ºN that brings it west (see the video of the 2500 m vorticity field in the Supplementary Material). The cyclone lifespan is between two and four weeks. The eddy first looses its strength moving south into deeper waters and then is dissipated by interactions with the bathymetry (see the vorticity video available as Supplementary Material).

Assuming that the lower layer is uniform and its potential vorticity (PV) is conserved, and that the relative vorticity atop the Fan is close to zero at times preceding the cyclone generation when local currents are weak, the strength of the cyclones can be evaluated considering the topographic stretching induced by the Fan, following Greene et al. (2009). The Fan extends between ~ 2700 and 1800 m, while the cyclonic PV anomaly extends upward to ~ 1000 m once the eddies begin drifting away from it. The vorticity generated by stretching is therefore approximately

$$\zeta = \left(\frac{H_{PVanom}}{H_{atop}} - 1\right) f = \left(\frac{2700 - 1000}{1800 - 1000} - 1\right) f \approx 1.1 f$$

The modeled relative vorticity at the cyclone cores in HR is indeed $1 \leq \frac{|\zeta|}{f} \leq 1.3$ around 2700 m, depending on the event considered, and decreases to about 0.8 ± 0.1 at 1500 m. At the time of formation, the cyclone intensity depends on resolution and in LR the modeled Fan eddies are weaker than in HR with maximum vorticity $|\zeta|/f \leq 0.9$. The vertical extent of the deep eddies is approximately consistent with a scale H ~ fL/N, where L is the topographic feature width and N is the sub-pycnocline buoyancy frequency.

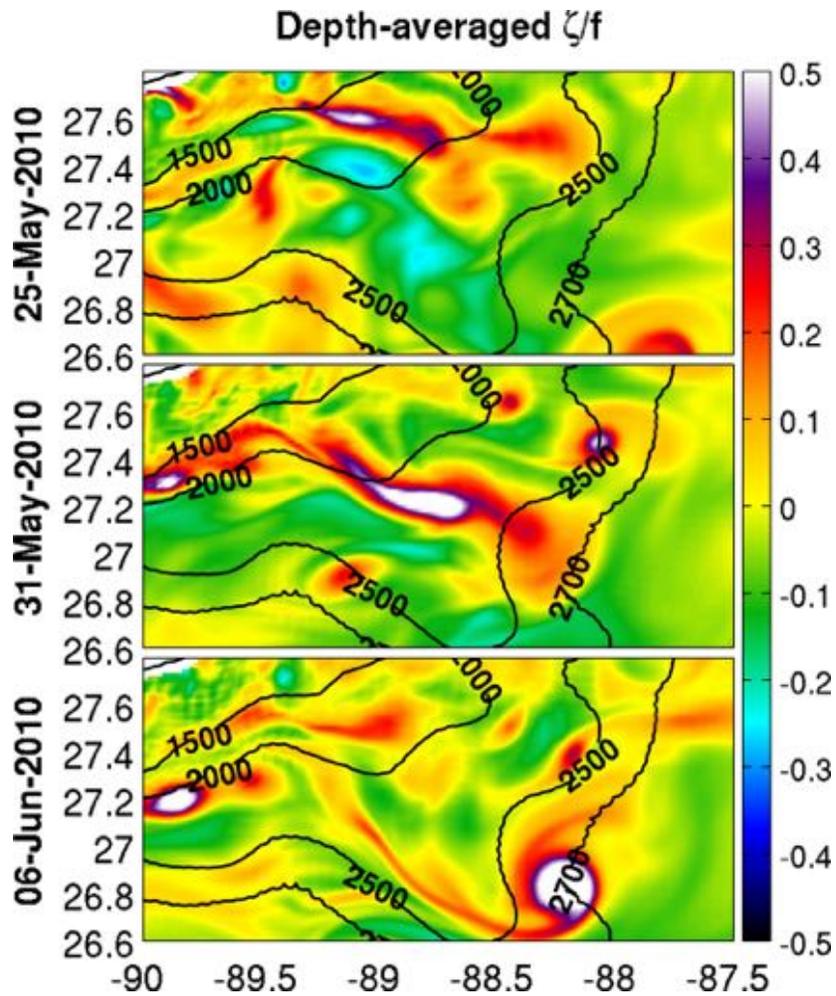

**Figure 5.** Snapshots of depth-averaged relative vorticity at the Mississippi Fan during the formation of one of the mesoscale cyclones in the HR run.

Owing to the resolution-dependent characteristics of these flow structures that dominate the lower layer vorticity, the domain averaged $\zeta/f$ is more positive in HR than LR below 1000 m, as shown in Fig. 6 (left panel). In the upper 1000 m, the anticyclonic vorticity of the Loop Current and Eddies dominate the domain averaged vorticity budget. In HR, however, the LC contribution is partially compensated in the mixed layer by submesoscale cyclonic eddies (see Fig. 6 in LBCM).

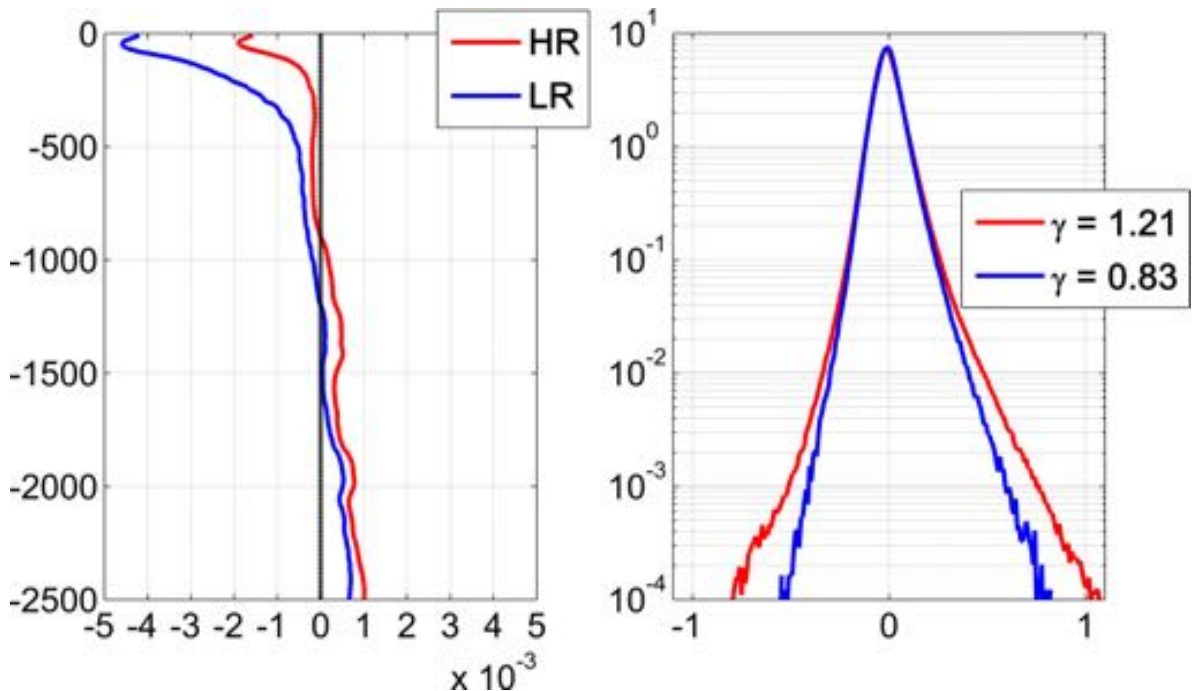

**Figure 6.** Left: Vertical profiles of relative vorticity $\zeta$ averaged over the northern Gulf domain in HR and LR; Right: PDFs of $\zeta/f$ in HR and LR at the depth of 1000 m, normalized to have unit integral probability. The skewness $\gamma$ of the two distributions is listed.

Many cylonic and anticyclonic eddies, smaller than the ones formed over the Fan, occupy the continental slope (Fig. 4). At 1000 m the shape of the PDFs of $\zeta/f$ (Fig. 6, right panel) indicates that both cyclonic and anticyclonic structures contribute to the tails of the distributions, and that model resolution plays an important role in the representation of those features. Below 2000 m fewer eddies beside the Fan cyclone can be found at both resolutions, and those formed are mostly cyclonic (with an approximate ratio of 3 cyclones to 1 anticyclone), so the tails of the PDFs are more skewed towards positive values (not shown). These smaller eddies increase in number and strength approaching the slope independent of resolution, but they are more intense and numerous in HR, in agreement with the exploration of resolution dependence in the representation of the California Undercurrent (Molemaker et al., 2015).

The eddies and filaments found along the GoM continental slope are generated through instabilities of horizontal shear layers generated near the bottom, where steep density gradients can develop, by two mechanisms. Firstly, those layers can be associated with the presence of strong and highly variable, zonal, front-like currents, and are formed at the front boundary, as in the case of the cyclonic structure at 26.2°N in Fig. 7. The submesoscale cyclone is formed in presence of a negative $\frac{\partial u}{\partial y}$ gradient, for the zonal velocity varies from nearly zero to highly negative for increasing $y$. Zonal velocities along the northern Gulf of Mexico slope are highly intermittent in direction and strength as observed in mooring data (Hamilton, 2007; Kolodziejczyk et al., 2012). This is confirmed in Fig. 7 and further in Fig. 8 through maps of near-bottom speed and its standard deviation (STD) at the two resolutions. The mean near-bottom velocity at or below 1000 m is westward over most of the northern Gulf (including over the Fan, while the anticyclonic recirculation appears at approximately 50 m above the bottom) and similar in pattern and direction in HR and LR. The speed of near bottom currents, $\left(\sqrt{u^2 + v^2}\right)$, however, is greater in HR. In both runs the STD is larger than the mean value, and bottom currents often reverse their direction. The STD also increases from LR to HR, with the consequent generation of a larger number of both cyclones and anticyclones in HR.

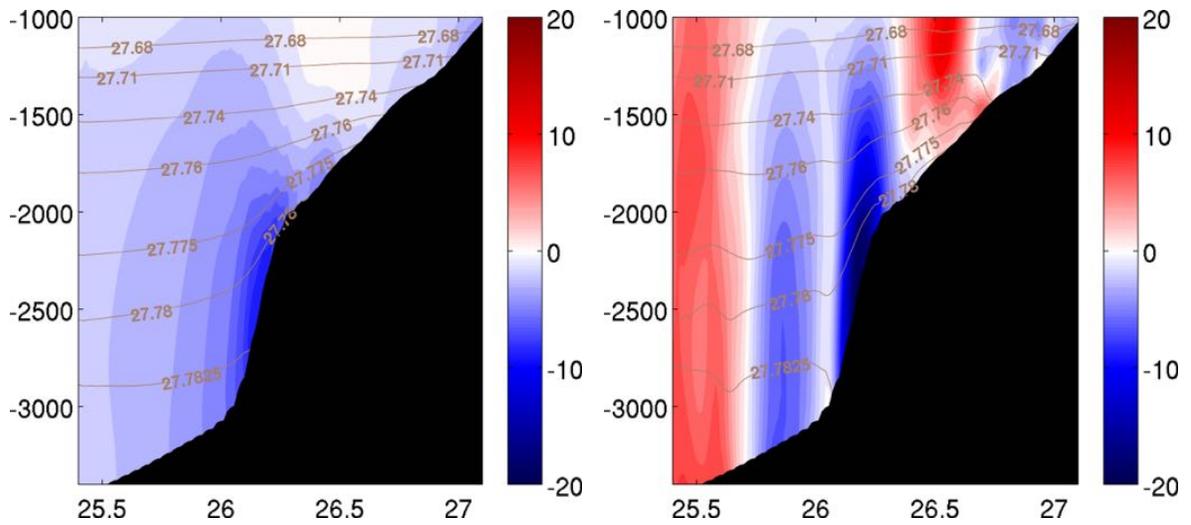

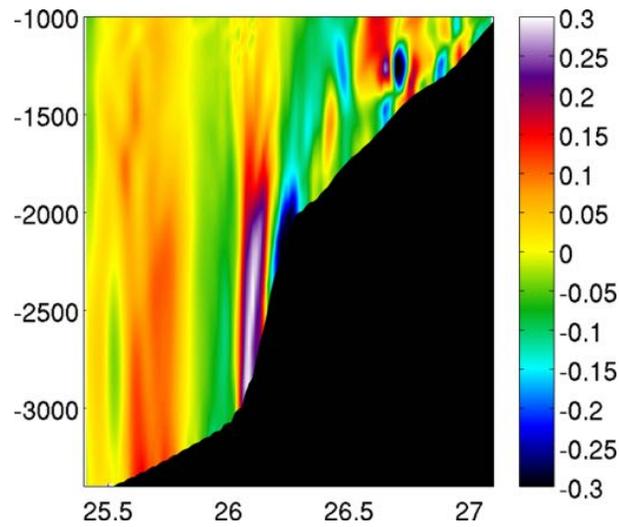

Figure 7. Mean (a) and instantaneous snapshot (b) of zonal velocity in cm s$^{-1}$, and the corresponding instantaneous relative vorticity normalized by the Coriolis frequency (c) along the -93.5°W transect in the HR run. Density isolines (mean or instantaneous) are superimposed on the velocity plots.

Secondly, shear layers are formed by the mechanism of current bottom drag on a slope acting within the bottom boundary layer, as proposed by Molemaker et al. (2015). In this case, the horizontal shear layer results from the bottom boundary layer parameterization employed by ROMS-Agrif (KPP, Large et al., 1994) that exerts a drag on the lowest model layers proportional to the square of the near bottom velocity (see schematic in Figure 7 in Molemaker et al. (2015)). Again, the shear layer supports the generation of vertical vorticity $\zeta \approx -\frac{\partial u}{\partial y}$ at the slope, where $u$ has to approach zero, and the sign of the along-slope current corresponds to the sign of $\zeta$ at the bottom boundary layer. Whenever high values of vorticity are achieved ($|\zeta|/f > 0.2$), small scales eddies and filaments are generated often in association with boundary current separation. This mechanism is exemplified in Fig. 7 by the anticyclonic eddy centered in the middle of the zonal 'jet', at 26.4°N, in correspondence of the core of the jet-like zonal current. In this case, the horizontal boundary layer width is proportional to the boundary layer vertical thickness and inversely proportional to the cross-shelf bottom slope (Molemaker et al., 2015), which is highly variable in the domain considered but often as steep as 0.25 between 1000 and 2700 m, despite the 5 km smoothing applied to both solutions. Therefore, the horizontal boundary layer is partially resolved in HR and mostly not resolved by LR. Preferred locations for the formation of the along-slope eddies can be found in Fig. 9, where vorticity maxima and minima are shown at 1000 m depth in both runs, and in Fig. 10, where the variance of the horizontal divergence field, $\frac{\partial u}{\partial x} + \frac{\partial v}{\partial y}$, normalized by the Coriolis frequency quantifies the departure from geostrophic balance. Anticyclones dominate the vorticity maxima at the Northwest

Slope and in the De Soto Canyon, while cyclones apart from the Fan eddies are preferentially formed along the Mississippi and Central Slopes, and at the Northwest Slope. Figure 9 confirms the LR underestimation of both strength and number of eddies generated along the continental slope. Finally the divergence field (Fig. 10) supports relatively large vertical velocities for the flows along the continental slope and atop major topographic features, which is indicative of ageostropic dynamics of appreciable amplitude.

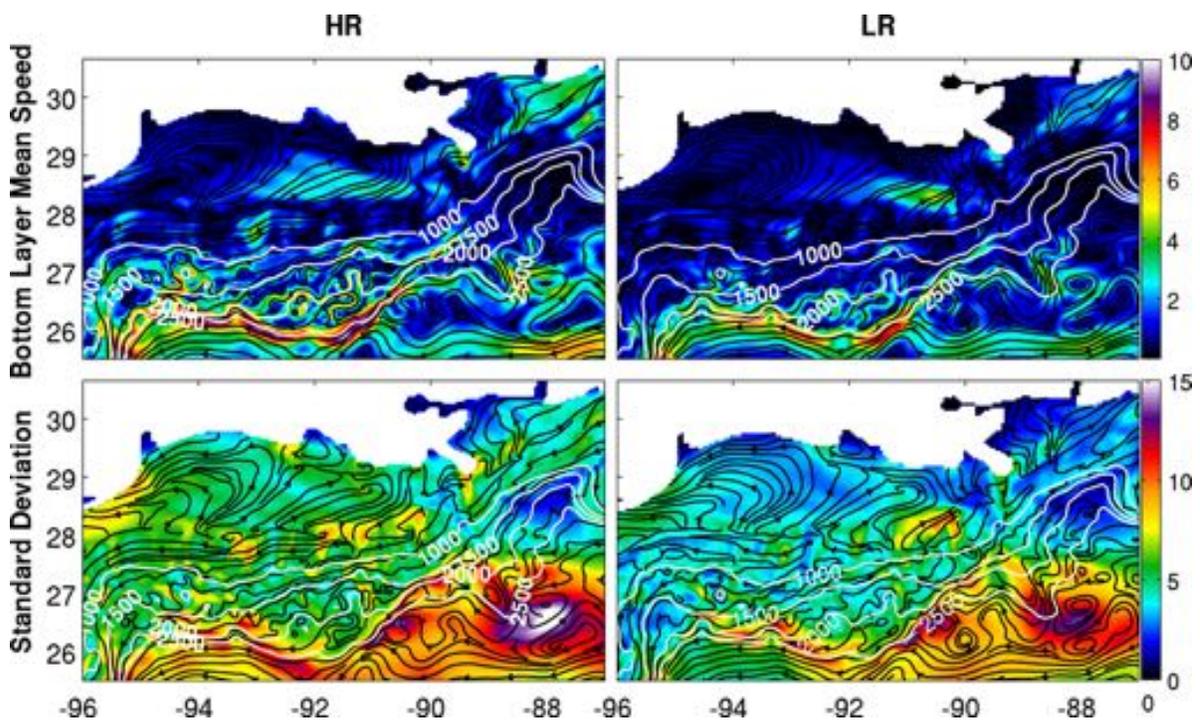

**Figure 8.** Mean near-bottom speed and its standard deviation at each model grid point in HR (left) and LR (right) with superimposed horizontal velocity streamlines in the bottom model layer. Unit: cm s$^{-1}$

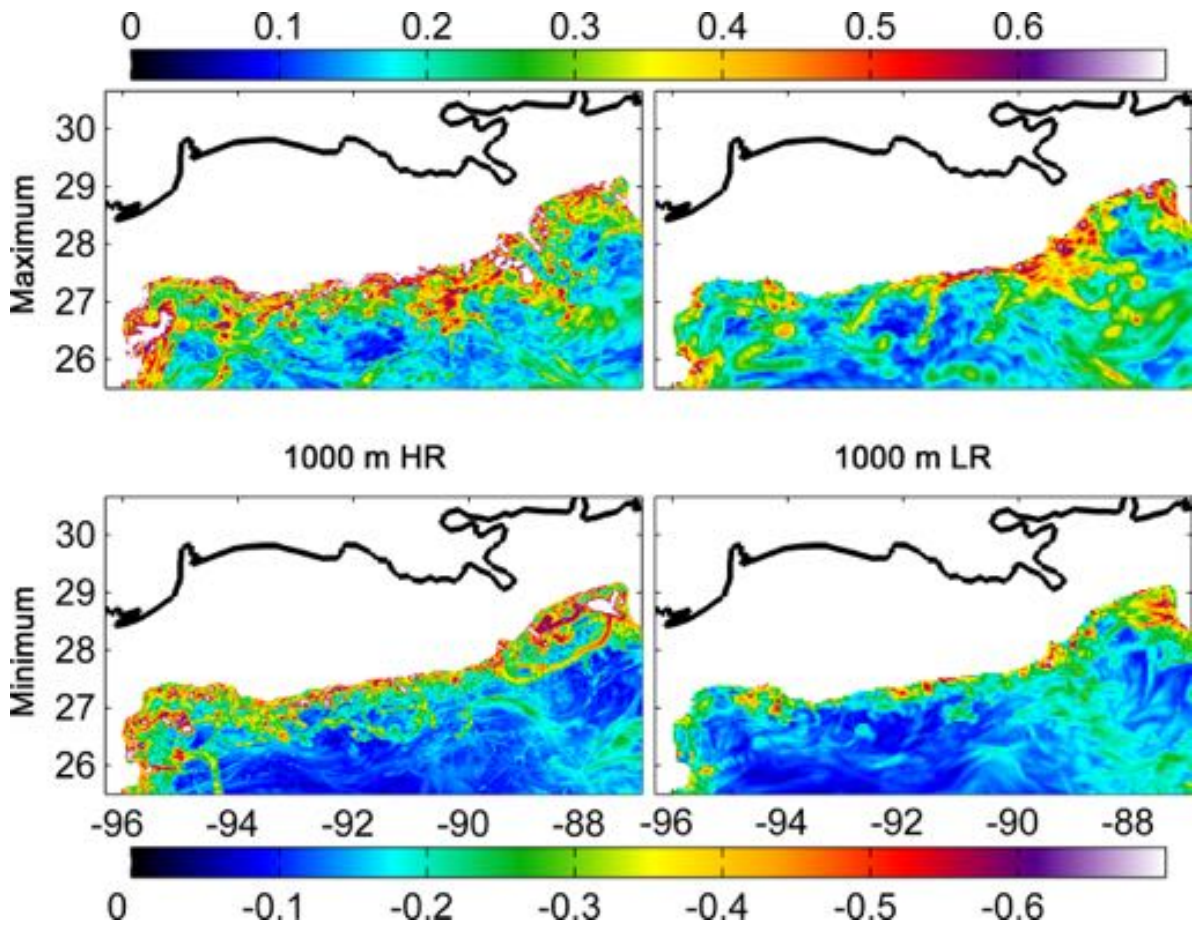

**Figure 9.** Maxima (top) and minima (bottom) of $\zeta/f$ at each model grid point in HR (left) and LR (right) at the depth of 1000 m. The association with the continental slope is strong.

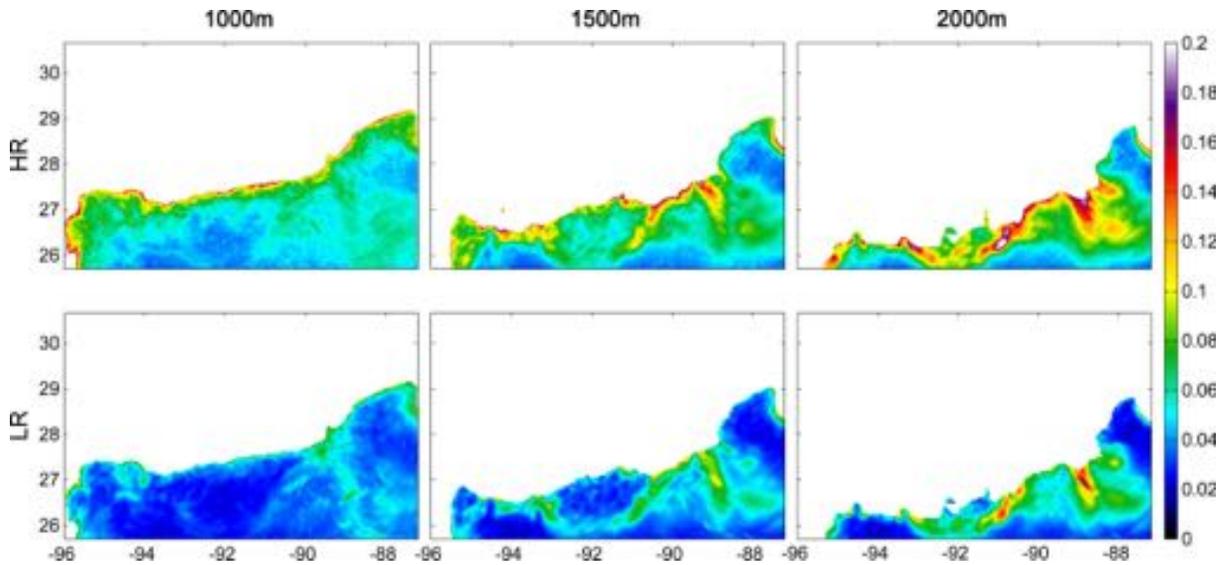

**Figure 10** Standard deviation of horizontal divergence normalized by the Coriolis frequency at 1000, 1500 and 2000 m in HR (top) and LR (bottom).

## 5. Passive tracer releases

Passive tracers are released in the bottom model layer, over approximately the same volume in the two simulations near the DWH site, at MC297, and along the eastern edge of the Central Slope, at GC600 (Fig. 1). The tracer advection and diffusion scheme adopted (Marchesiello et al., 2009; Lemarié et al., 2012) does not guarantee positive concentrations, although it does greatly diminish false extrema and diapycnal tracer flux[2]. Negative values appear in correspondence of large tracer gradients and they are retained to evaluate the impact of resolution on the scheme accuracy. The LR simulation is impacted more dramatically: in the first month after deployment 15% of the concentration values in the LR simulation are negative. Such percentage decreases to 5% or less past the first trimester of tracer integration. In the HR case percentages are limited to 5% in the first month, and decrease to 1%.

---

[2] Alternative advection schemes avoid negative concentrations, but they are generally more erroneously diffusive. We prefer to accept the error levels shown here to retain the weakly diffusive behavior of the advection operator in ROMS.

The two deployment sites differ significantly in their circulation. Around the MC site, the near-bottom mean circulation consists of very weak cyclonic currents (less than 1 cm s$^{-1}$) with the lowest standard deviation in the deep domain but still sufficient to reverse the mean direction at times (Fig. 8). Sporadic, intense submesoscale eddies of both signs form along the continental slope and populate the Mississippi and De Soto Canyon areas in HR, while weaker, less coherent and shorter living structures are found in the LR solution (see videos of the vorticity field at 1000 and 1500 m in the Supplementary Material). At GC600 HR and LR differ mostly in the representation of the mean along-slope cyclonic currents to the east and west of the site, that reaches 5 cm s$^{-1}$ at HR but at most 2.5 cm s$^{-1}$ in LR, and of their variance, underestimated in LR. The circulation in the immediate proximity of GC600 is affected by topographically constrained mesoscale eddies that form along the Mississippi slope at the site' depth independently of resolution but are generally more intense in HR, and are advected towards the Central Slope (Fig. 4 and vorticity field video at 1500 m). Over the Central Slope vorticity shows little coherency at about the GC600 depth (~1250 m) independently of resolution due to the complex bathymetry that quickly erode the submesoscale eddies formed in HR near the bottom at depths of 1700-2200 m, as highlighted by the vorticity videos.

In the case of the MC tracer, the along- and across-slope spreading is initially more confined in HR, due to the improved resolution of the meridional extension of the along-slope currents and to the trapping exerted by the submesoscale eddies formed in the Canyons (Fig. 11).

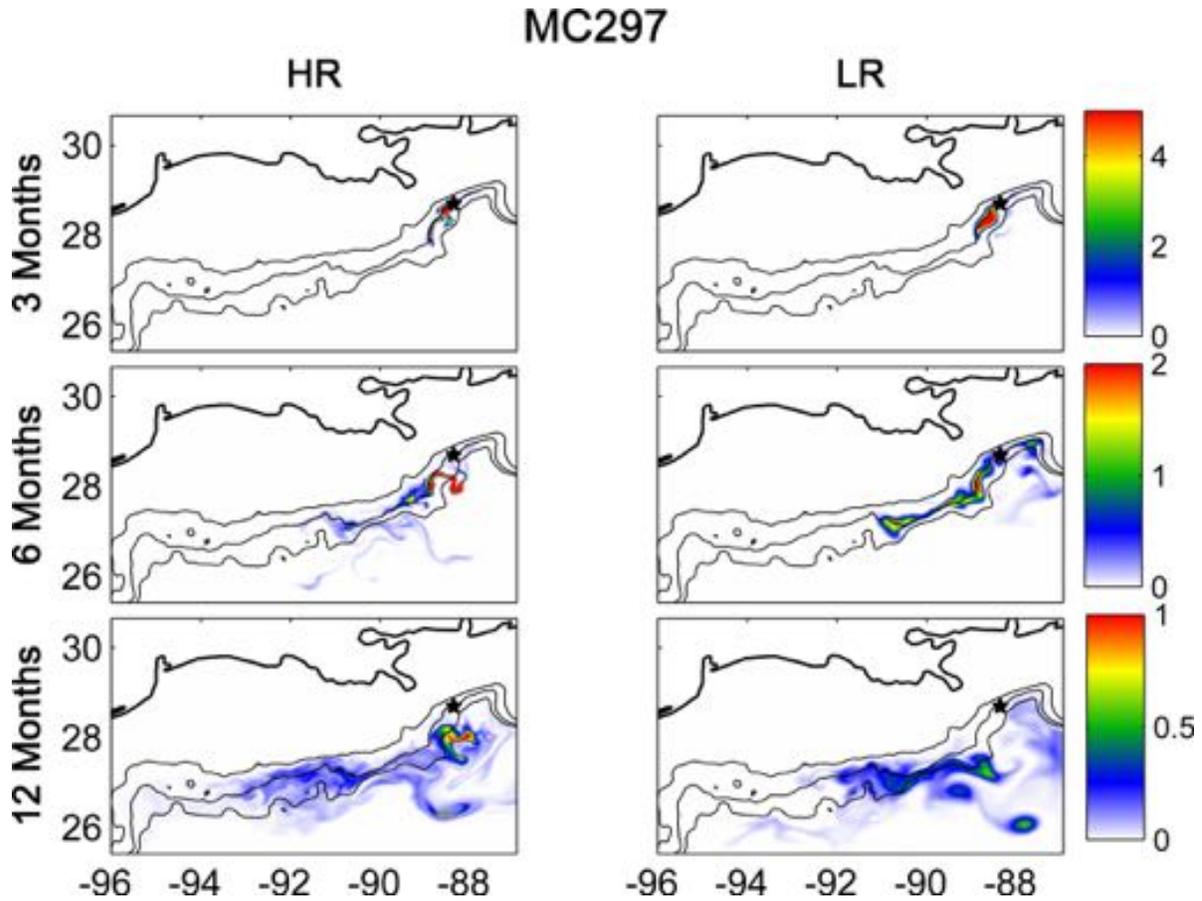

**Figure 11** Depth-integrated concentration of the tracer deployed at MC297 (black star) after 3, 6 and 12 months after deployment in the HR (left) and LR (right) integrations. The 1000, 1500 and 2000 m isobaths are also shown.

In the HR run three months after deployment most of the MC297 tracer can be found in the immediate proximity of the site, in a narrow 100 km long filament to the west of it and, in very small amount, in a filament the east, extending into the De Soto Canyon along the 1500 m isobath. The tracer to the west follows very closely the trajectory of the deep oil plume identified in June 2010 (Camilli et al., 2010) matching well the plume zonal and meridional extension. During the 2010 spill, small negative oxygen anomalies, indicative of microbial oil consumption, were also recorded in the De Soto area (http://www.noaa.gov/deepwaterhorizon/maps/dissolved_maps.html). Six months after deployment large concentrations of the MC297 tracer are still found within 75 km of the site, while some tracer has spread west of the Mississippi mouth carried by the along-slope currents and submesoscale eddies. The patchiness and overall extension of the westward propagating tracer is associated with advection by a chain of submesoscale eddies that form regularly along the 1500 m isobath to the west of the Mississippi Fan in the HR case (see Fig. 4 and vorticity movies). Comparable patchiness and distribution was found in the post-spill dissolved oxygen negative anomalies integrated between 1000 and 1300 m and July 22$^{nd}$ and October 16$^{th}$ 2010 (http://www.nodc.noaa.gov/Images/profile_rost_chartLarge.jpg). One year after deployment the HR tracer has occupied most of the area comprised between the 1500 and 2000 m isobaths, but it is not uniformly distributed and a significant percentage (~10%) of tracer is still concentrated to the immediate south and south-west of the deployment site. Such spreading is broadly consistent with the limited footprint of *DWH* oil deposited on the ocean floor by the deep-water plume (Valentine et al., 2014). In LR the MC297 tracer after three months has spread into a blob and moved to the west, and after six months is distributed further to the west, along a narrow band of uniform concentration that follows the bathymetry contours as far as -91$^{o}$W. Only in the last six months mesoscale dynamics, and particularly the cyclones formed at the Mississippi Fan introduce patchiness in the tracer distribution (see videos of the evolution of the tracer concentration from 3 to 12 months after deployment in HR and LR in the Supplementary Material).

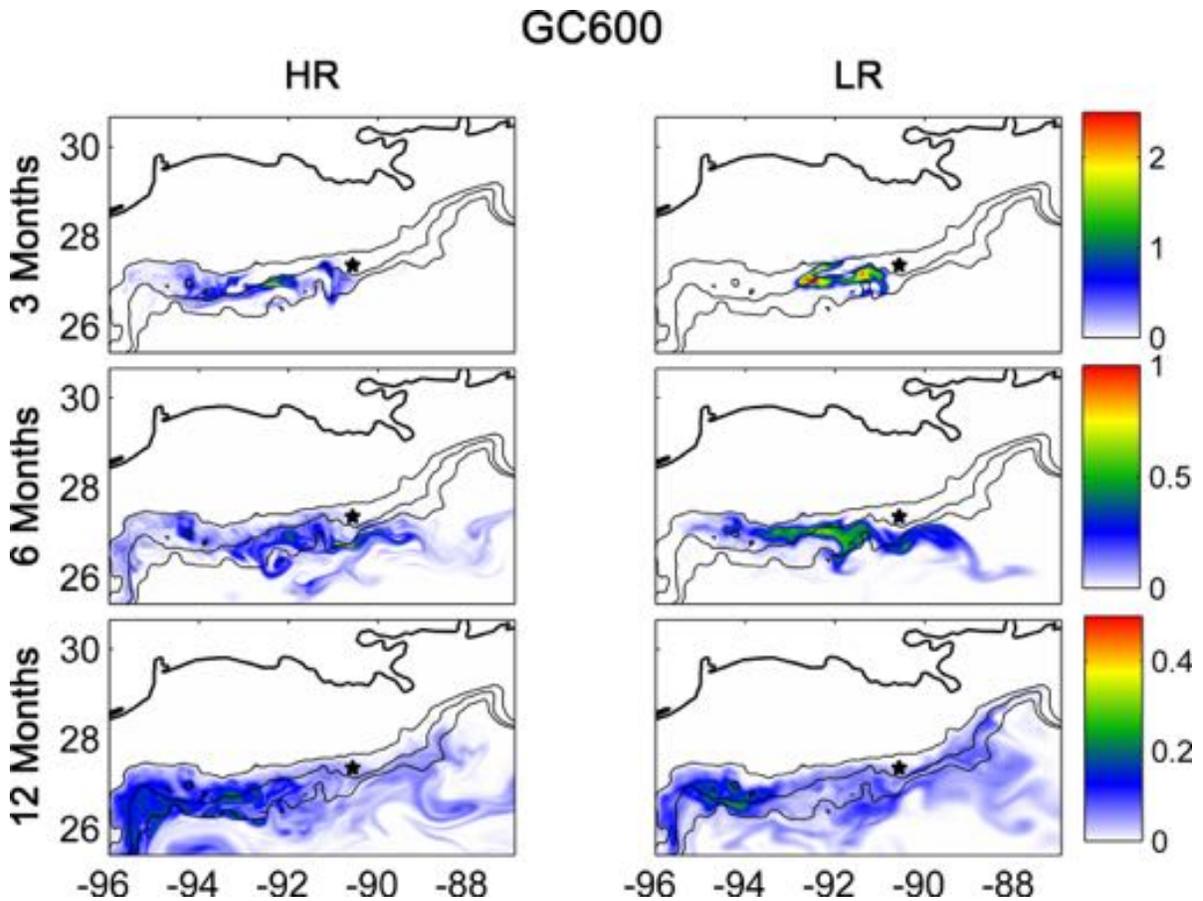

**Figure 12** Depth-integrated concentration of the tracer deployed at 600 (black star) after 3 (top), 6 (middle) and 12 (bottom) months after deployment in the HR (left) and LR (right) integrations. The 1000, 1500 and 2000 m isobaths are also shown. Note the different color scale compared to Fig. 11.

A different picture emerges at GC600 (Fig. 12). Here the near bottom velocity around the site plays an important role in the tracer evolution over the first few months. In both runs the tracer follows a comparable trajectory towards the Central slope, but it is much faster in HR due to the stronger velocities. Once over the slope the tracer is quickly diffused, while continuing in its propagation mostly, but not exclusively, to the west. At later times, once the tracer has covered most of the slope, mixing appears to be slightly more efficient in the LR case, due to the limited number of trapping eddies. After one year the GC600 tracer can be found in small concentrations over most of the domain in waters deeper than 1000 m, with the exception of the De Soto Canyon.

To quantify differences and similarities in tracer patchiness and lateral spreading efficiency in the two releases, Table 1 presents the cumulative percentage of the total amount of tracer contained in the 250 km² with highest concentration in both runs.

| Date | MC297 HR | MC297 LR | GC600 HR | GC600 LR |
| --- | --- | --- | --- | --- |
| 07/20/2010 | 45% | 27% | 4% | 7.5% |
| 10/20/2010 | 15.5% | 7% | 2% | 2% |
| 1/20/2011 | 6.5% | 5.5% | 0.8% | 1.3% |
| 4/20/2011 | 3% | 1.7% | 0.5% | 0.9% |

**Table 1:** Cumulatively percentage of total tracer contained in the 250 km² with strongest concentration 3, 6, 9 and 12 months after release.

We estimated the tracer-based horizontal (meridional and lateral) diffusivities from depth-integrated concentrations, C(x,y). Our estimates are based on the growth of the second moment of the tracer cloud under the assumption that once the velocities are decorrelated the growth is linear in time (Taylor, 1921; Garrett, 1983), following previous applications (Ledwell et al., 1993; Klocker et al., 2012). For example, the zonal diffusivity component is given by

$$K_{x,t} = \frac{1}{2}\frac{\partial}{\partial t}\frac{\iint (x-x_b)^2 C(x,y)dxdy}{\iint C(x,y)dx\,dy}$$

where C is interpolated with a volume weight to conserve the total mass over the whole Gulf domain and $x_b$ is the x-coordinate of the tracer cloud center of mass

$$x_b = \frac{\iint xC(x,y)\,dx\,dy}{\iint C(x,y)\,dx\,dy}.$$

The diffusivities are shown in Fig. 13 and are calculated retaining the negative concentrations (bold lines) or setting them to zero at each time step (thin lines) to highlight the uncertainties associated with numerical noise. Uncertainties are negligible in HR but considerable and across all times in LR due to the large negative values generated in the first few weeks of simulation. The second order moments of the tracer cloud evolve similarly in HR and LR in the MC297 deployment, and so the diffusivities, despite the different patchiness. Diffusivities are higher in the HR run around GC600 due to the fast spread of the dye in the second month of integration, but level off after about six month, while they continue to grow in both zonal and meridional direction in LR. The asymptotic value of $K_x$ in the GC600 deployments is twice as large the MC297 case, while $K_y$ is independent of the site in the HR run. $K_y$ is 5 to 10 times smaller than $K_x$ and none of the curves follow a simple diffusive behavior (ballistic and/or Richardson-like).

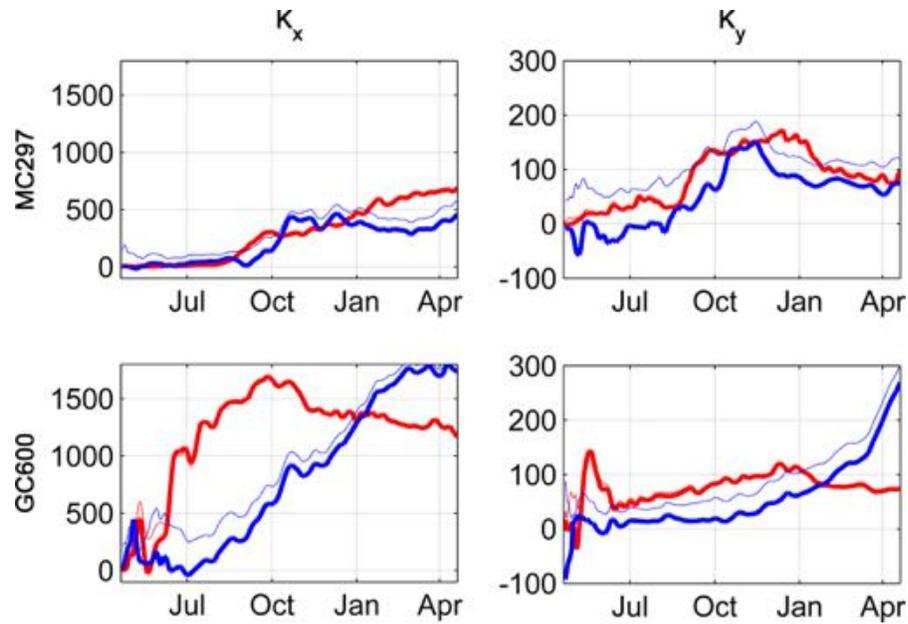

**Figure 13** Zonal ($K_x$) and meridional ($K_y$) diffusivities as function of time in HR (red) and LR (blue), including concentrations that include negative values (bold) or after removing them. Note the different y-axis for $K_x$ and $K_y$. Unit: $m^2\,s^{-1}$.

The variability in K(t) and the patchiness of the spatial patterns in Figs. 11-12 show considerable sampling fluctuations compared to ensemble-mean statistics expected from multiple realizations. While acknowledging this limitation, we have chosen to restrict our attention here to a single realization for each tracer release due to its relevance to interpret episodic pollutant releases, as in the case of the 2010 spill.

Finally, we evaluated how the vertical distribution of the tracers is affected by resolution. In the MC deployment differences between the two runs are large in the first six months, when the tracers in HR is trapped in bottom intensified submesoscale filaments and eddies, and spread less effectively through the water column than in LR. After one year, the patch of tracer confined around the deployment site in HR has moved to the south reaching greater depths than in the LR case (Fig. 14). Modest differences in vertical spreading are found in the GC600 deployment, due to the absence of long lasting trapping features over the Central slope at depths of 1500 m or less (not shown).

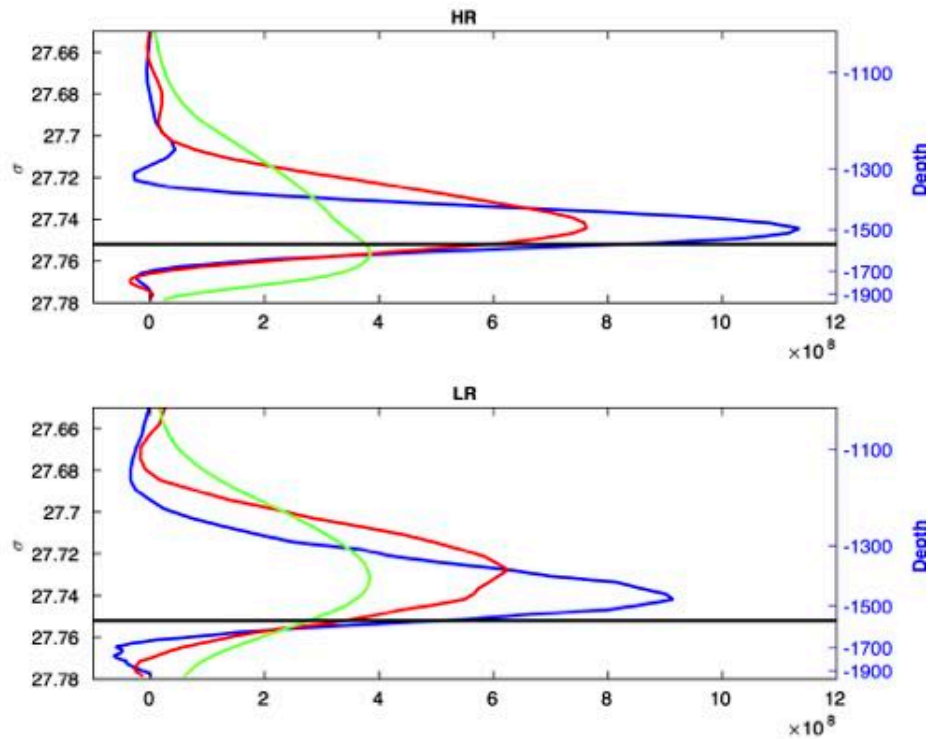

**Figure 14** Vertical profiles of horizontally averaged tracer mass (concentration multiplied by the volume occupied) in density space 3 (blue), 6 (red) and 12 (green) months after deployment at MC297 transformed into depth z through the 3-year mean density profile calculated over an area covering the tracer 6 months after deployment. The black line indicates the release depth. Depth unit: m; $\sigma = \rho - 1000$ kg m$^{-3}$, integrated mass of the tracer: $10^{10}$.

We estimated the standard deviation and vertical diffusivity of pairs of profiles separate by a time Δt using a Gaussian fit. The diapycnal diffusivity $k$ is given by $k = \frac{1}{2}\left(\sigma_f^2 - \sigma_i^2\right)/\Delta t$ where $\sigma_f^2$ and $\sigma_i^2$ are the standard deviation of the vertical spread obtained from the Gaussian fit at the final and initial time (here taken as 6 days).

|        | HR MC297 | LR MC297 | HR GC600 | LR GC600 |
|--------|----------|----------|----------|----------|
| 23-May | 0.7      | 1.7      | 3.7      | 3.5      |
| 25-Jun | 0.7      | 2.1      | 2.9      | 1.5      |
| 20-Jul | 0.4      | 0.9      | 2.2      | 2.2      |
| 20-Oct | 0.6      | 0.8      | 2.1      | 1.2      |
| 20-Jan | 0.6      | 0.9      | 1.5      | 1        |
| 20-Apr | 0.5      | 0.8      | 1.1      | 0.9      |

**Table 2:** Diapycnal diffusivity coefficients, $k$, calculated from the standard deviation of Guassian fits to the tracer profiles in $10^{-3}$ m² s$^{-1}$.

While the vertical spreading behaviors in Fig. 14 have a plausible pattern, the diffusivity values in Table 2 are somewhat higher than usually measured by microstructure and dye release (in particular during the deep release experiment conducted in 2012 in the vicinity of MC297; Jim Ledwell, personal communication). Because discrete advection errors can cause false diapycnal tracer flux in ROMS, we cannot say how much this might be contaminating our $k$ estimates. Nevertheless, submesoscale currents are known to influence energy dissipation and thus diapycnal flux, and the estimated $k$ values are at least suggestive that they might be important over slope regions.

## 6. Conclusions

The occurrence and impact of submesoscale currents below 1000 m in the northern Gulf of Mexico has been investigated with two three-year long regional simulations differing only in their horizontal resolution. In the first integration dx = 1.6 km (HR) and the submesoscale processes are partially resolved, while in the second one dx = 5 km (LR) and the submesoscales are mostly unresolved. A combination of mesoscale and submesoscale eddies are generated over the continental slope. The largest eddy in both integrations is a bottom-intensified cyclone with a diameter of about 70 km. It forms by vortex stretching whenever the circulation in the anticyclonic recirculation zone over the Mississippi Slope is sufficiently strong to advect the flow over the Mississippi Fan. The submesoscale eddies are (partially) resolved only in the HR run, are not in geostrophic balance, and are formed through instabilities of bottom shear layers. Those layers are generated by the juxtaposition of along-slope frontal currents that are highly variable in both speed and direction, and more so for increasing resolution, or by current bottom drag on the continental slope. Whenever formed by bottom drag, the shear boundary layer is in most cases under-resolved in the low resolution run being its width proportional to the slope of the bathymetry. Submesoscale cyclones and anticyclones are in approximately equal number between 1000 and 1500 m, while cyclones dominate below. They form everywhere along the continental slopes, as shown in Fig. 9, and are dissipated preferentially over the Central Slope and Sigsbee Escarpment through interactions with the complex bathymetry. Submesoscale eddies generated in the De Soto Canyon live the longest, up to four months, being isolated from strong shear currents, while the numerous structures that populate the Northwest Slope are usually sheared away within a couple weeks.

Submesoscale eddies and vorticity filaments, and the more intense and variable currents found in the HR run impact the representation of transport and mixing along the slope in ways that vary greatly across the northern GoM, as shown by the passive tracer integrations. We deployed a tracer close to the Macondo wellhead, at MC297, and a second one to the west of the Mississippi Slope, at GC600. In the MC297 site, the mean flow velocity around the deployment site is small. In HR the tracer is trapped inside submesoscale eddies and stretched into narrow filaments, resulting in a patchy distribution, while in LR it spreads in a uniform narrow band that develops along the continental slope. The mapping of the deep oil plume following the 2010 spill supports the patchiness found in HR. Near the GC600 site, on the other hand, the mean currents are strong, and their time and space variability is large and increases with resolution. In the higher resolution run the tracer is transported faster from the site to the Central Slope, where is efficiently diffused due to the turbulent interaction of the bottom flow with the complex bathymetry.

The two tracer experiments exemplify the complexity and variability of the circulation along the continental slope in the northern Gulf of Mexico. Overall, our results indicate that the predictability potential in the event of another deep spill is hampered by mesoscale and submesoscale structures that form at irregular intervals along most of the continental slope, and by the variability of the along-slope currents. While the long-term, asymptotic behavior of pollutants originating from different sites in the Gulf may be estimated (general direction of propagation, minimum and maximum times required to be advected and diffused from one location to another), the trapping in individual eddies and filaments is not. In the aftermath of the 2010 spill those submesoscale 'patches' of hydrocarbons and dispersants were likely responsible for the spotty impacts to deep coral communities (Fisher et al., 2014).

A potential delicate issue is the extent of the numerical error introduced in tracer integrations in presence of strong concentration gradients. The HR simulation indicates that the error is small when the horizontal resolution is sufficiently high, but it is large enough to impact diffusivity estimates in a mesoscale-resolving integration.

Finally, this investigation confirms that the generation of submesoscale eddies along continental slopes is a generic process, as hypothesized by Molemaker et al. (2015), likely relevant to most coastal areas, with important implications for understanding abyss-to-coast connectivity.

## Acknowledgment

This work was made possible by a grant (in part) from BP/the Gulf of Mexico Research Initiative to support consortium research entitled "Ecosystem Impacts of Oil and Gas Inputs to the Gulf (ECOGIG)" administered by the University of Georgia - Athens. JCM acknowledges support from the Office of Naval Research, grant N00014-12-1-0939.. The modeling data relevant to this work may be obtained through the Gulf of Mexico Research Initiative Information and Data Cooperative (GRIIDC) under number:

R1.x132.141:0005 or, in the case of complete 3D fields, contacting the first author. ECOGIG contribution number: ??.

**Appendix: Validation of near bottom model velocities**

The mean circulation of the Gulf of Mexico, in both the observations and in ROMS is discussed in Cardona and Bracco (2014) for a similar configuration. A validation of the model surface currents and of the temperature and salinity distributions across the water column in the current set-up is presented in the Appendix of LBCM. Overall ROMS provides a very good representation of all mean quantities, including the Loop Current (LC) strength and evolution, the Loop Eddies detachment and propagation, the Yucatan Channel transport, mean near surface and deep currents, and the water column stratification.

The mean deep flow in the northern Gulf is not well characterized in the observations due to the paucity of data. A mean velocity reconstruction over this area has been attempted using 17 PALACE (Profiling Autonomous Lagrangian Circulation Explorer) floats deployed at approximately 900 m between 1998 and 2002 (Weatherly et al., 2004). At this depth the flow is still dominated, both in the model and in the observations, by the Loop Current. Nonetheless the velocity field derived from float trajectories supports the modeled mean cyclonic pattern, the presence of recirculation zones, and the mean speed amplitude presented in Fig. 2.

As mentioned, topographic Rossby waves with periodicities from 10 to 50 days, contribute to the variability of the deep flow of the northern Gulf (Hamilton, 2009; Kolodziejczyk et al., 2012; Pérez-Brunius et al., 2013). In the observations the variability associated to TRWs has been shown to have periodicities preferentially in the 10-20 days band for currents aligned to the 1000-1500 m isobaths, 20-30 days for currents along the 1500 - 2000 m isobaths, and around 50 days for near-bottom currents over 2300 – 2500 m (Hamilton, 2009). We verified that comparable periodicities are simulated by ROMS, despite the smoothing applied to the bathymetry, in both solutions. This is shown in Fig. A1 (HR case only), where we plot the period of the maximum of the variance-preserving spectrum for the first normalized empirical orthogonal function (EOF-1) of the zonal velocity $u$ for each grid point at which the variance is higher than an arbitrary threshold, and over two four-months long periods at the beginning and end of the integration (similar results are obtained through other model periods). The threshold was chosen based on the observed speed spectra (Hamilton, 2009) and set to be 0.5.

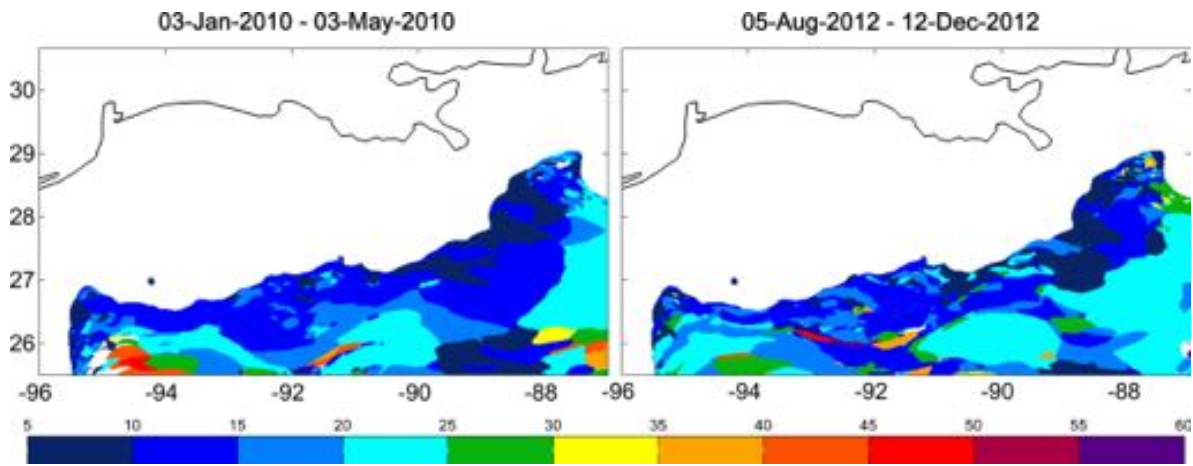

**Figure A1** Maps of the period of the maximum of the variance-preserving spectrum for the normalized EOF-1 of the near-bottom zonal velocity $u$ whenever the variance is higher than 0.5 in the HR solution. Left panel: January-May 2010; Right panel: August-December 2012. Unit: Days.

Additionally, numerical simulations have shown that in the north GoM the eddy kinetic energy (EKE) to kinetic energy (KE) ratio is close to one everywhere but over the portion of the continental shelf where the water column is shallower than 200 m (Cardona and Bracco, 2014). In the northern Gulf mesoscale dynamics prevails also in the bottom layer, contributing at least 80% of the total kinetic energy.

Deep current measurements in our focus area are not abundant. They are mostly limited to the eastern end of the Central Slope, where the topography is particularly steep, with current meters deployed as early as the 1980s (Hamilton, 1990; Hamilton, 2007), and to exploratory work performed between 2003 and 2006 (Donohue et al., 2006; Hamilton, 2009). The magnitude of the modeled currents at eastern portion of the Central Slope underestimates the observations. A maximum speed of 90 cm s$^{-1}$, associated with the propagation of TRWs with a short period of 8-14 days, has been recorded at about 50 m from the bottom (Hamilton and Lugo-Fernandez, 2001). In contrast, modeled values attain only 50 cm s$^{-1}$ in HR and 30 cm s$^{-1}$ in LR, and the characteristic period over the area covering the mooring is 10-25 days. The speed underestimation is caused in part by the smoothing applied to the topographic relief to limit unphysical pressure gradient currents. We tested that values nearing 80 cm s$^{-1}$ are achieved near the bottom along the steeper portions of the continental close in a sensitivity run performed at 1 km resolution with the bathymetry smoothed at 1 km (in this work the bathymetry is smoothed at 5 km in both runs). Longer preferred periodicities, ranging from 40 to 80 days, characterize the modeled flow south of 26.5$^o$N in the eastern portion of the domain, in agreement with previous modeling work (Oey and Lee, 2002), while 20-40 day peaks are commonly found in spectra calculated over areas covering the slope to the west of the eastern end of Sigsbee. The periodicities are independent of resolution given that an identical smoothing is applied to the topography (not shown).

Between 2010 and 2013, the ECOGIG program collected several time series of near bottom velocities (usually 10 m from the bottom) through Acoustic Doppler Current Profilers (ADCPs) and single point current meters deployed on lander systems at the locations where the passive tracers have been released. The time series were obtained in correspondence of bathymetric features with size of few hundred meters that are not resolved by our model. Notwithstanding the modeled mean speed $\left(\sqrt{u^2+v^2}\right)$ compares reasonably well with the observed ones subsampled at the model frequency and over the same time period (MC297: ADCP =1.6 cm s$^{-1}$; HR = 2.8 cm s$^{-1}$; LR = 2.0 cm s$^{-1}$. GC600: ADCP 5.4 cm s$^{-1}$; HR = 4.1 cm s$^{-1}$; LR = 2.6 cm s$^{-1}$). Sample time-series of the meridional and zonal velocity components at the three sites on intervals overlapping with the available in-situ data are presented in Fig. A2.

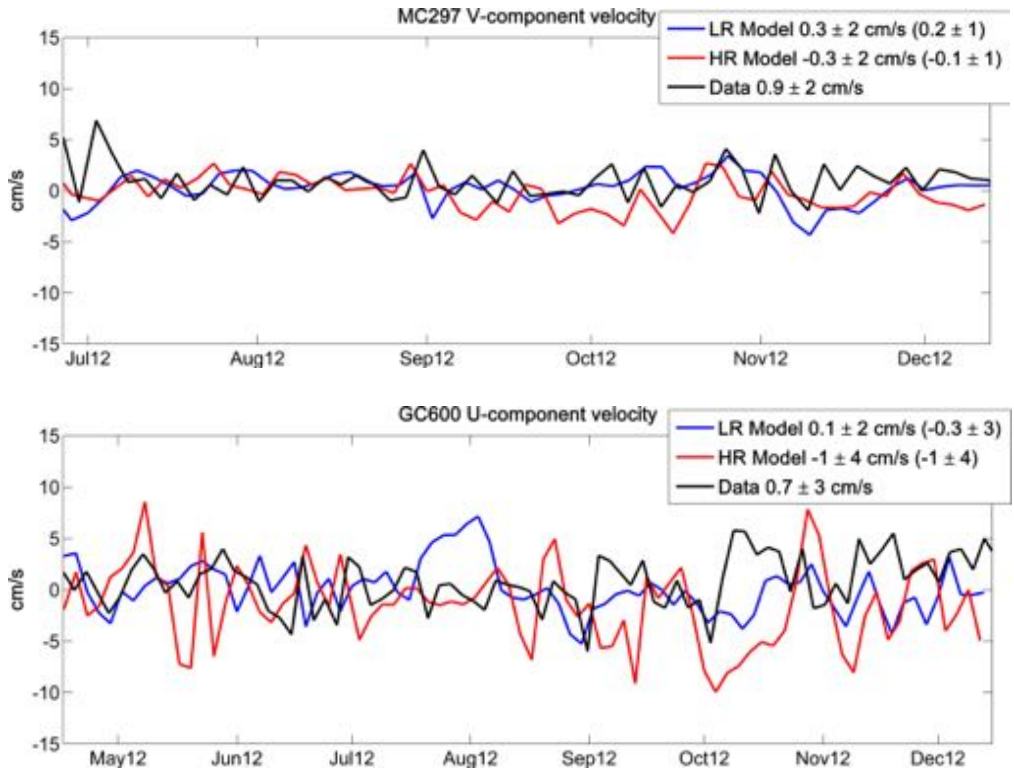

**Figure A2:** Sample time-series of 3-day averages of meridional (a) and zonal (b - c) velocity in the model and in-situ data at the MC297 and GC600 sites, respectively. Mean and standard deviation calculated over the observational time frame are indicated in the legend. For the model integrations mean and standard deviation over the Jan. 2010 - Dec. 2012 period are also shown in parenthesis. Unit: cm s$^{-1}$.

**Figure Captions**

**Figure 1.** Bathymetry over the northern Gulf (HR) domain. The 1000, 1500, and 2000 m bathymetric contours are indicated by the black lines and major topographic features are named. The sites where the passive tracer is released are indicated by black stars.

**Figure 2.** Mean velocity vector field in the whole Gulf of Mexico with zooms in the northern portion in both HR (top left) and LR (top right). The color bar indicates speed in cm s$^{-1}$.

**Figure 3.** Eddy Kinetic Energy time series averaged over the northern Gulf at different depths across the deep layer. HR run on top (thick lines) and LR integration below (thin lines). The black dots indicate the generation of mesoscale cyclones at the Mississippi Fan. Unit: cm$^2$s$^{-2}$

**Figure 4.** Snapshots of relative vorticity normalized by the Coriolis frequency, $\zeta/f$, depth-averaged from 1000 m to the bottom. Top panel: HR; Bottom panel: LR.

**Figure 5.** Snapshots of depth-averaged relative vorticity at the Mississippi Fan during the formation of one of the mesoscale cyclones in the HR run.

**Figure 6.** Left: Vertical profiles of relative vorticity $\zeta$ averaged over the northern Gulf domain in HR and LR; Right: PDFs of $\zeta/f$ in HR and LR at the depth of 1000 m,

normalized to have unit integral probability. The skewness γ of the two distributions is listed.

**Figure 7**. Mean (a) and instantaneous snapshot (b) of zonal velocity in cm s$^{-1}$, and the corresponding instantaneous relative vorticity normalized by the Coriolis frequency (c) along the -93.5°W transect in the HR run. Density isolines (mean or instantaneous) are superimposed on the velocity plots.

**Figure 8.** Mean near-bottom speed and its standard deviation at each model grid point in HR (left) and LR (right) with superimposed horizontal velocity streamlines in the bottom model layer. Unit: cm s$^{-1}$

**Figure 9.** Maxima (top) and minima (bottom) of $\zeta/f$ at each model grid point in HR (left) and LR (right) at the depth of 1000 m. The association with the continental slope is strong.

**Figure 10** Standard deviation of horizontal divergence normalized by the Coriolis frequency at 1000, 1500 and 2000 m in HR (top) and LR (bottom).

**Figure 11** Depth-integrated concentration of the tracer deployed at MC297 (black star) after 3, 6 and 12 months after deployment in the HR (left) and LR (right) integrations. The 1000, 1500 and 2000 m isobaths are also shown.

**Figure 12** Depth-integrated concentration of the tracer deployed at 600 (black star) after 3 (top), 6 (middle) and 12 (bottom) months after deployment in the HR (left) and LR (right) integrations. The 1000, 1500 and 2000 m isobaths are also shown. Note the different color scale compared to Fig. 11.

**Figure 13** Zonal ($K_x$) and meridional ($K_y$) diffusivities as function of time in HR (red) and LR (blue), including concentrations that include negative values (bold) or after removing them. Note the different y-axis for $K_x$ and $K_y$. Unit: $m^2 s^{-1}$.

**Figure 14** Vertical profiles of horizontally averaged tracer mass (concentration multiplied by the volume occupied) in density space 3 (blue), 6 (red) and 12 (green) months after deployment at MC297 transformed into depth z through the 3-year mean density profile calculated over an area covering the tracer 6 months after deployment. The black line indicates the release depth. Depth unit: m; $\sigma = \rho - 1000$ kg m$^{-3}$, integrated mass of the tracer: $10^{10}$.

**Figure A1** Maps of the period of the maximum of the variance-preserving spectrum for the normalized EOF-1 of the near-bottom zonal velocity *u* whenever the variance is higher than 0.5 in the HR solution. Left panel: January-May 2010; Right panel: August-December 2012. Unit: Days.

**Figure A2:** Sample time-series of 3-day averages of meridional (a) and zonal (b - c) velocity in the model and in-situ data at the MC297 and GC600 sites, respectively.

Mean and standard deviation calculated over the observational time frame are indicated in the legend. For the model integrations mean and standard deviation over the Jan. 2010 - Dec. 2012 period are also shown in parenthesis. Unit: cm s$^{-1}$.